\documentclass{article}
\usepackage{spconf}

\usepackage{amsmath}

\DeclareMathOperator*{\argmin}{arg\,min}
\usepackage{amsfonts}
\usepackage{amssymb}
\usepackage{dsfont}[sans]
\usepackage{graphicx}
\usepackage{cite}
\usepackage{bm}
\usepackage{enumitem}
\usepackage{cancel}
\usepackage{array}

\usepackage{tikz}
\usepackage{pgfplots}
\usetikzlibrary{arrows,matrix}
\usetikzlibrary{plotmarks}
\usepgfplotslibrary{patchplots}
\usepackage{grffile}

\usepackage{tabulary}
\usepackage{multirow}

\definecolor{mycolor1}{RGB}{36, 255, 36}   
\definecolor{mycolor2}{RGB}{0, 109, 219}   
\definecolor{mycolor3}{RGB}{146, 0, 0}     
\definecolor{mycolor4}{RGB}{219, 209, 0}   
\definecolor{mycolor5}{RGB}{211,211,211}   

\newcounter{tempEquationCounter}
\newcounter{thisEquationNumber}

\usepackage{rotating}
\usepackage{makecell}
\usepackage{colortbl}
\usepackage{arydshln}

\usepackage{amsmath,amsfonts}
\usepackage{bm}
\usepackage{cite}
\usepackage[bbgreekl]{mathbbol}

\title{GEVD-based Low-Rank Channel Covariance Matrix Estimation and MMSE Channel Estimation for Uplink Cellular Massive MIMO Systems}

\name{Robbe~Van~Rompaey, Marc~Moonen
    \thanks{The work of R. Van Rompaey was supported by a doctoral Fellowship of the Research Foundation Flanders (FWO-Vlaanderen). This research work was carried out at the ESAT Laboratory of KU Leuven, in the frame of Research Council KU Leuven: C24/16/019 ``Distributed Digital Signal Processing for Ad-hoc Wireless Local Area Audio Networking" and Fonds de la Recherche Scientifique - FNRS and Fonds Wetenschappelijk Onderzoek - Vlaanderen EOS Project no 30452698 `(MUSE-WINET) MUlti-SErvice WIreless NETwork'. The scientific responsibility is assumed by its authors. }}
\address{KU Leuven\\
    Dept. of Electrical Engineering-ESAT, STADIUS\\
    Kasteelpark Arenberg 10, B-3001 Leuven, Belgium}

\begin{document}
\ninept
\maketitle
\begin{abstract}
Uplink channel estimation is a crucial component for the performance of cellular massive MIMO systems. However, when the number of user equipments (UEs) grows, the sharing of the available resources causes interference between UEs in neighboring cells. Minimum mean squared error (MMSE) channel estimators have been proposed to mitigate this interference, but these require estimates of the channel covariance matrices. Therefore, a new channel covariance matrix estimator for low-rank channel covariance matrices is presented in this paper, using a generalized eigenvalue decomposition (GEVD) of two covariance matrices that can be estimated from the available uplink data. The requirements for the systems are minimal and, except for synchronization, there is no need for communication between the different cells and no prior knowledge on the background noise is required. Approximate MMSE estimators are also derived based on the newly proposed channel covariance matrix estimator. The effectiveness of the proposed methods is demonstrated in numerical simulations.
\end{abstract}
\begin{keywords}
Covariance matrix estimation, random pilot allocation, low-rank channel model, generalized eigenvalue decomposition (GEVD), massive MIMO
\end{keywords}
\section{Introduction}

Massive MIMO (multiple input multiple output) is a key technology for current generation cellular networks \cite{Hoydis2013,Bjornson2018,Bjornson2019a}. Base stations (BSs) are deploying hundreds of antennas, enabling spatial multiplexing of several user equipments (UEs) per cell, both in uplink and downlink, with simple linear signal processing methods \cite{Rusek2013}. The increasing number of UEs, and as a consequence the sharing of the available pilots whenever the coherence blocks are limited in size, causes interference between UEs in neighboring cells. To make it possible for a BS to separate the UEs that it serves from the interference, second-order statistical information present in the channel covariance matrices has to be exploited during the channel estimation.

The channel covariance matrices are commonly assumed to be perfectly known in massive MIMO literature, which is a strong assumption since the matrix dimension grows with the number of antennas and furthermore the statistics are changing over time. Existing covariance matrix estimators either use additional pilot overhead, sacrifice samples for data transmission \cite{Bjornson2017a}, use strong separability conditions that might not always hold \cite{Haghighatshoar2017}  or require knowledge of the used pilots of all the UEs in the network \cite{Neumann2018, Xie2016}. Therefore, a new channel covariance matrix estimator for low-rank channel covariance matrices is presented in this paper, using a generalized eigenvalue decomposition (GEVD) of two covariance matrices that can be estimated from the available uplink data. The requirements for the system are minimal and, except for synchronization, there is no need for communication between the different cells and no prior knowledge on the background noise is required. Approximate MMSE estimators are also derived based on the newly proposed channel covariance matrix estimator. The effectiveness of the proposed methods is demonstrated in numerical simulations.

\section{Channel and Signal Model}
\label{cemMIMO:sec:channel_signal}

\subsection{Channel Model}
A cellular massive MIMO system with $L$ cells is considered where each cell contains a BS with $N$ antennas and $K$ single-antenna UEs. The channel from UE $k$ in cell $l$ to the BS in cell $j$ is denoted by $\mathbf{h}_{jlk} \in \mathbb{C}^N$  and is assumed to remain constant during a coherence block of $\tau_c$ samples. Superscript $.^t$ will be used to denote the quantities in coherence block $t$, e.g.  $\mathbf{h}^t_{jkl}$. A common channel model for $\mathbf{h}^t_{jkl}$ is that it is drawn from a correlated Rayleigh fading realization $\mathcal{NC}(\mathbf{0},\overline{\mathbf{R}}_{jlk})$, where $\overline{\mathbf{R}}_{jlk} \in \mathbb{C}^{N \times N}$ is the positive semi-definite channel covariance matrix describing the large-scale fading, including geometric pathloss, shadowing, antenna gains, and spatial channel correlation \cite{Bjornson2017}. The complex Gaussian distribution around $\overline{\mathbf{R}}_{jlk}$ models the small-scale fading. Due to the different positions of the UEs, it is assumed that the channels for different UEs are uncorrelated, i.e. $E\{\mathbf{h}^t_{jlk}  \mathbf{h}^{t,H}_{jmi}\} = \mathbf{0}$ if $\{l,k\} \ne \{m,i\}$, where $E\{.\} $ denotes the expected value operator, with respect to different channel realizations and $.^H$ is the Hermitian transpose operator. Measurements in \cite{Viering2002} show that the matrices $\{\overline{\mathbf{R}}_{jlk}\}$ remain constant over several thousands of coherence blocks, where this large number is denoted by $T$.

The scattering is mostly localized around the UE, since the BSs are elevated and hence have limited scattering in their near-field \cite{Xie2016}. The multipath components defining the total channel $\mathbf{h}_{jlk}$ are therefor arriving from a specific localized region in the network, so that the matrices $\{ \overline{\mathbf{R}}_{jlk} \}$ can be approximated by low-rank matrices, where the rank is denoted by $R$.  $\overline{\mathbf{R}}_{jlk}$ can thus be approximated as a sum of $R$ rank-1 terms:
\begin{equation}
\label{cemMIMO:eq:R_rank}
\overline{\mathbf{R}}_{jlk} = \sum_{r=1}^{R} \overline{\mathbf{q}}_{jlk,r} \overline{\mathbf{q}}_{jlk,r}^H
\end{equation}
where $\overline{\mathbf{q}}_{jlk,r}$ will be defined later. This implies that the channel $\mathbf{h}^t_{jlk}$ in coherence block $t$ can be modeled  as
\begin{equation}
\label{cemMIMO:eq:model_h}
\mathbf{h}^t_{jlk} = \sum_{r=1}^{R} \overline{\mathbf{q}}_{jlk,r} \overline{z}^t_{jlk,r} = \overline{\bm{\mathds{Q}}}_{jlk} \overline{\mathbf{z}}_{jlk}^t
\end{equation}
where the normal complex variables $\overline{\mathbf{z}}_{jlk}^t = [\overline{z}^t_{jlk,1} \ ... \ \overline{z}^t_{jlk,R}]^T$ with $.^T$ the transpose operator, are independently drawn from $\mathcal{NC}(0,1)$ and $\overline{\bm{\mathds{Q}}}_{jlk} = [\overline{\mathbf{q}}_{jlk,1} \ ... \ \overline{\mathbf{q}}_{jlk,R}]$.

\subsection{Signal Model}
The samples in one coherence block of $\tau_c$ samples are divided in $\tau_p$ samples for uplink channel estimation and $\tau_u$ samples for uplink data transmission with $\tau_c = \tau_p + \tau_u$. With  $\tau_p$ samples available, the network can predefine $\tau_p$ different orthogonal and unitary pilot sequences $\{\mathbf{s}_b\}$ where
\begin{equation}
\mathbf{s}_b = [s_b(1) \ ... \ s_b(\tau_p)]^T, \ b=1...\tau_p
\end{equation}
with $\mathbf{s}^H_b \mathbf{s}_b = \tau_p$ and $\mathbf{s}^H_b \mathbf{s}_c = 0$ if $b \ne c$. In the uplink channel estimation phase, each UE $k$ in each cell $l$ selects randomly one of the $\tau_p$ pilot sequences in each coherence block $t$ , the index of this pilot sequence is denoted by $b^t_{lk}$. An example is provided in Table \ref{cemMIMO:tab:training_sequences}. The UEs transmits the chosen pilot sequence with powers $\{p_{lk}\}$. The signal received at BS $j$ is then given as
\begin{equation}
\label{cemMIMO:eq:y_l-training}
\mathbf{y}^t_{j}(p) = \sum_{l=1}^L \sum_{k=1}^K \sqrt{p_{lk}} \mathbf{h}^t_{jlk} s_{b_{lk}^t}(p) +  \mathbf{n}^t_j(p), \ p=1...\tau_p
\end{equation}
where $p$ is the sample index and $\mathbf{n}^t_j(p)$ is the background noise at BS $j$. 

\begin{table}
    \hspace*{-0.16in}
    \resizebox{1.04\columnwidth}{!}{
    \begin{tabular}{l|c|c:c:c:c|c:c:c:c|c:c:c:c|cccc|c:c:c:c|}
        \multicolumn{1}{c}{} & \multicolumn{1}{c}{} & \multicolumn{20}{c}{{\large Coherence block}} \\ 
        \cline{3-22}
        \multicolumn{1}{c}{\begin{sideways}\end{sideways}} &  & \multicolumn{4}{c|}{} & \multicolumn{4}{c|}{} & \multicolumn{4}{c|}{} & \multicolumn{4}{c|}{} & \multicolumn{4}{c|}{} \\ [-1em]
        \multicolumn{1}{c}{\begin{sideways}\end{sideways}} &  & \multicolumn{4}{c|}{{\large 1}} & \multicolumn{4}{c|}{{\large 2}} & \multicolumn{4}{c|}{{\large 3}} & \multicolumn{4}{c|}{...} & \multicolumn{4}{c|}{{\large $T$}} \\ 
        \cline{3-22}
        \multicolumn{1}{c}{\begin{sideways}\end{sideways}} &  & $\mathbf{s}_1$ & $\mathbf{s}_2$ & $\mathbf{s}_3$ & $\mathbf{s}_4$ & $\mathbf{s}_1$ & $\mathbf{s}_2$ & $\mathbf{s}_3$ & $\mathbf{s}_4$ & $\mathbf{s}_1$ & $\mathbf{s}_2$ & $\mathbf{s}_3$ & $\mathbf{s}_4$ & \multicolumn{4}{c|}{...} & $\mathbf{s}_1$ & $\mathbf{s}_2$ & $\mathbf{s}_3$ & $\mathbf{s}_4$ \\ 
        \cline{2-22}
        \multirow{5}{*}{\rotcell{{\large UE}}} & 1 &  & {\cellcolor[rgb]{0.753,0.753,0.753}} &  &  &  &  & {\cellcolor[rgb]{0.753,0.753,0.753}} &  & {\cellcolor[rgb]{0.753,0.753,0.753}} &  &  &  &  & \multicolumn{2}{c}{...} & &  &  &  & {\cellcolor[rgb]{0.753,0.753,0.753}} \\ 
        \cline{2-22}
        & 2 &  &  & {\cellcolor[rgb]{0.753,0.753,0.753}} &  & {\cellcolor[rgb]{0.753,0.753,0.753}} &  &  &  &  &  & {\cellcolor[rgb]{0.753,0.753,0.753}} &  &  \multicolumn{4}{c|}{...}  &  & {\cellcolor[rgb]{0.753,0.753,0.753}} &  &  \\ 
        \cline{2-22}
        & 3 & {\cellcolor[rgb]{0.753,0.753,0.753}} &  &  &  &  &  & {\cellcolor[rgb]{0.753,0.753,0.753}} &  &  &  & {\cellcolor[rgb]{0.753,0.753,0.753}} &  &  \multicolumn{4}{c|}{...}  &  &  & {\cellcolor[rgb]{0.753,0.753,0.753}} &  \\ 
        \cline{2-22}
        & 4 &  &  &  & {\cellcolor[rgb]{0.753,0.753,0.753}} & {\cellcolor[rgb]{0.753,0.753,0.753}} &  &  & & {\cellcolor[rgb]{0.753,0.753,0.753}}   &  &  &  &  \multicolumn{4}{c|}{...}  & {\cellcolor[rgb]{0.753,0.753,0.753}}  &  &  &  \\ 
        \cline{2-22}
        & ... & \multicolumn{1}{c}{} & \multicolumn{2}{c}{...} &  & \multicolumn{1}{c}{} & \multicolumn{2}{c}{...} &  & \multicolumn{1}{c}{} & \multicolumn{2}{c}{...} &  &  \multicolumn{4}{c|}{...}  & \multicolumn{1}{c}{} & \multicolumn{2}{c}{...} &  \\ 
        \cline{2-22}
        & $K$ &  & {\cellcolor[rgb]{0.753,0.753,0.753}} &  &  &  &  &  & {\cellcolor[rgb]{0.753,0.753,0.753}} & {\cellcolor[rgb]{0.753,0.753,0.753}} &  &  &  &  \multicolumn{4}{c|}{...}  &  &  & {\cellcolor[rgb]{0.753,0.753,0.753}} &  \\
        \cline{2-22}
    \end{tabular}
    }
    \caption{Example of random allocation of 4 pilot sequences over $K$ UEs in a certain cell.}
    \label{cemMIMO:tab:training_sequences}
\end{table}

In the uplink data transmission phase, each UE $k$ in each cell $l$ transmits the unitary signal $s_{lk}(u)$ with zero mean and power $p_{lk}$. The received signal at BS $j$ is then given as
\begin{equation}
\label{cemMIMO:eq:y_l-data}
\mathbf{y}^t_{j}(u) = \sum_{l=1}^{L} \sum_{k=1}^K \sqrt{p_{lk}} \mathbf{h}^t_{jlk} s_{lk}(u) + \mathbf{n}^t_j(u), \ u=1...\tau_u.
\end{equation}
Note that the indices $p$ and $u$ are used to distinguish between the different phases. BS $j$ has to detect the uplink data from the UEs in its cell, i.e. $s_{jk}(u)$ for $k=1...K$. To perform receive combining strategies like maximum-ratio (MR), (regularized) zero-forcing (RZF) or minimum mean squared error (MMSE) combining \cite{Bjornson2017,Ozdogan2019}, good estimates of the channels of the $K$ UEs in cell $j$ are required. The background noise $\mathbf{n}_j^t$ is assumed to be zero-mean complex Gaussian, but with an unknown spatial covariance matrix $\overline{\mathbf{R}}_{\mathbf{n}_j\mathbf{n}_j} = E \{ \mathbf{n}_j^t \mathbf{n}_j^{t,H}\}$. It is often assumed to be white noise with a diagonal covariance matrix, but here a more general noise model is used, where the noise can be correlated between the antennas of a BS. This happens for example when there are located jamming signals nearby the BS.

\section{Optimal MMSE Channel Estimation}
\label{cemMIMO:sect:optimal_MMSE}

The signal obtained from \eqref{cemMIMO:eq:y_l-training} after despreading with pilot sequence $b^t_{jk}$ can be represented as
\begin{align}
\label{cemMIMO:eq:y}
\mathbf{y}_{j}^{t,\text{pilot}}[b^t_{jk}] =& \sum_{u=1}^{\tau_p} \mathbf{y}^t_{j}(p) s^*_{b_{jk}^t}(p)  \\
= & \sum_{l=1}^L \sum_{i=1}^K \delta^{t}_{jk,li} \sqrt{p_{li}} \tau_p \mathbf{h}^t_{jli} + \underbrace{\sum_{u=1}^{\tau_p} \mathbf{n}^t_{j}(p) s^*_{b_{jk}^t}(p)}_{\mathbf{n}_{j}^{t,\text{pilot}}[b^t_{jk}]} \nonumber
\end{align}
where the random variable $\delta^{t}_{jk,li}$ is 1 with probability $1/\tau_p$ and 0 otherwise since all other UEs randomly pick 1 of the $\tau_p$ pilot sequences and  $\delta^{t}_{jk,jk}$ is deterministic and equal to 1.

Based on the despread signal in \eqref{cemMIMO:eq:y} a linear MMSE channel estimation criterion, similar to \cite[Sect. 4.2]{Demir;2021}, can be defined to obtain an estimate $\overline{\mathbf{h}}^t_{jjk} = \overline{\mathbf{W}}^H_{jk}\mathbf{y}_{j}^{t,\text{pilot}}[b^t_{jk}]$ of $\mathbf{h}_{jjk}^t$  in the current coherence block $t$:
\begin{equation}
\label{cemMIMO:eq:W_k-optimal}
\overline{\mathbf{W}}_{jk} = \argmin_{\boldsymbol{W}_{jk}} E \{ ||\mathbf{h}^t_{jjk} - \boldsymbol{W}^H_{jk} \mathbf{y}_{j}^{t,\text{pilot}}[b^t_{jk}]||^2 \}.
\end{equation}
The optimal solution of \eqref{cemMIMO:eq:W_k-optimal} is given as 
\begin{equation}
\label{cemMIMO:eq:W}
\overline{\mathbf{W}}_{jk} = \sqrt{p_{jk}} \left( \overline{\mathbf{R}}^{\text{pilot}}_{jjk} \right)^{-1} \overline{\mathbf{R}}_{jjk}
\end{equation}
where
\begin{align}
\label{cemMIMO:eq:R_k_dot}
\overline{\mathbf{R}}^{\text{pilot}}_{jjk} &= \frac{1}{\tau_p} E \{\mathbf{y}_{j}^{t,\text{pilot}}[b^t_{jk}] \mathbf{y}_{j}^{t,\text{pilot},H}[b^t_{jk}] \}  \\
&= p_{jk} \tau_p \overline{\mathbf{R}}_{jjk} + \sum_{i \ne k} p_{ji} \overline{\mathbf{R}}_{jji} + \sum_{l\ne j} \sum_{i=1}^K p_{li} \overline{\mathbf{R}}_{jli} + \overline{\mathbf{R}}_{\mathbf{n}_{j}\mathbf{n}_{j}} \nonumber
\end{align}
since the channels between different UEs are uncorrelated and are also uncorrelated with the background noise. The fact that $E \{ {\delta^{t^2}_{jk,li}}\} = \frac{1}{\tau_p}$ is also used.  For the background noise term, the following derivation is used:
\begin{align}
E\{&\mathbf{n}_{j}^{t,\text{pilot}}[b^t_{jk}] \mathbf{n}_{j}^{t,\text{pilot},H}[b^t_{jk}]\} \\
&= E\left\lbrace \left(\sum_{u=1}^{\tau_p} \mathbf{n}_{j}^t(p) s^*_{b_{jk}^t}(p)\right) \left(\sum_{u=1}^{\tau_p} \mathbf{n}_{j}^{t,H}(p) s_{b_{jk}^t}(p)\right)^H \right\rbrace \nonumber \\
&=  \sum_{u=1}^{\tau_p} E\{ |s_{b_{jk}^t}(p)|^2 \}  E\{\mathbf{n}_{j}^t \mathbf{n}_{j}^{t,H}\} = \tau_p \overline{\mathbf{R}}_{\mathbf{n}_{j}\mathbf{n}_{j}} \nonumber
\end{align} 
where the cross terms in the second step are zero since the noise is uncorrelated between different samples.

An important difference with the MMSE channel estimation method proposed in \cite[Sect. 4.1,]{Demir;2021} is that now all UEs appear as interfering UEs in \eqref{cemMIMO:eq:R_k_dot}, while in \cite{Demir;2021} it is assumed that the pilot sequences for all the UEs are predetermined and fixed, so that only UEs sharing the same pilot sequence appear as interfering UEs. However, this assumption requires prior knowledge of the channel statistics of all the UEs and the background noise, while the proposed method allows for a more efficient estimation method which will be discussed next. 

\section{Low-rank Channel Estimation}
\subsection{Low-rank covariance matrix estimator}
\label{cemMIMO:subsect:covariance_estimation}
In practice the covariance matrices $\overline{\mathbf{R}}^{\text{pilot}}_{jjk}$ and $\overline{\mathbf{R}}_{jjk}$ have to be estimated from the available data. $\overline{\mathbf{R}}^{\text{pilot}}_{jjk}$ can be estimated using time averaging over the last $T$ coherence blocks, possibly using a diagonal regularizer as in \cite{Ledoit2004,Bjornson2017a}, since perfect synchronization and knowledge of $b_{jk}^t$ at the BS $j$ is assumed. The estimation will be denoted as $\mathbf{R}^{\text{pilot}}_{jjk}$ (without the overline). $\overline{\mathbf{R}}_{jjk}$ as in \eqref{cemMIMO:eq:R_rank} can not immediately be estimated, since the channel is unknown. However, the combined covariance matrix $\mathbf{R}^{\text{all}}_{j}$ can be estimated from all available antenna signals during both the channel estimation phase and uplink data transmission phase, which is an estimate of
\begin{align}
\label{cemMIMO:eq:R}
\overline{\mathbf{R}}^{\text{all}}_{j} &=  E \left\lbrace \frac{1}{\tau_p} \sum_{p=1}^{\tau_p} \mathbf{y}_{j}^{t}(p) \mathbf{y}_{j}^{t,H}(p) + \frac{1}{\tau_u} \sum_{u=1}^{\tau_u} \mathbf{y}_{j}^{t}(u) \mathbf{y}_{j}^{t,H}(u) \right\rbrace  \nonumber \\
&= \sum_{l=1}^L \sum_{i=1}^K p_{li} \overline{\mathbf{R}}_{jli} + \overline{\mathbf{R}}_{\mathbf{n}_{j}\mathbf{n}_{j}} .
\end{align}
When comparing \eqref{cemMIMO:eq:R_k_dot} and \eqref{cemMIMO:eq:R}, it can be stated that $\overline{\mathbf{R}}^{\text{pilot}}_{jjk}$ is the covariance matrix estimated after despreading and $\overline{\mathbf{R}}^{\text{all}}_{j}$ is the covariance matrix estimated before despreading. They differ only in a contribution depending on $\overline{\mathbf{R}}_{jjk}$. The following relation can thus be derived:
\begin{equation}
\label{cemMIMO:eq:R_estimate}
p_{jk} \overline{\mathbf{R}}_{jjk} = \frac{1}{\tau_p-1} \left( \overline{\mathbf{R}}^{\text{pilot}}_{jjk} - \overline{\mathbf{R}}^{\text{all}}_{j} \right).
\end{equation}
However, simply replacing $\overline{\mathbf{R}}^{\text{pilot}}_{jjk}$ and $\overline{\mathbf{R}}^{\text{all}}_{j}$ with the estimated quantities $\mathbf{R}^{\text{pilot}}_{jjk}$ and $\mathbf{R}^{\text{all}}_{j}$ in \eqref{cemMIMO:eq:R_estimate} will not result in a good estimate $\mathbf{R}_{jjk}$ of $\overline{\mathbf{R}}_{jjk}$ since it will generally not satisfy the assumed rank assumption in \eqref{cemMIMO:eq:R_rank} and might even result in an indefinite estimate $\mathbf{R}_{jjk}$. One known remedy consists in using a rank-approximation based on the GEVD \cite{Serizel2014, Dendrinos1991} of the matrix pencil $\{{\mathbf{R}}^{\text{pilot}}_{jjk}, {\mathbf{R}}^{\text{all}}_{j} \}$ given as
\begin{align}
{\mathbf{R}}^{\text{pilot}}_{jjk} &= \mathbf{Q}_{jjk} \bm{\Sigma}_{jjk} \mathbf{Q}^H_{jjk} = \sum_{r=1}^{N} \sigma_{jjk,r} \mathbf{q}_{jjk,r} \mathbf{q}^H_{jjk,r}\\
{\mathbf{R}}^{\text{all}}_{j} &= \mathbf{Q}_{jjk} \mathbf{Q}^H_{jjk} = \sum_{r=1}^{N} \mathbf{q}_{jjk,r} \mathbf{q}^H_{jjk,r}
\end{align}
where $\bm{\Sigma}_{jjk} = \text{diag}\{\sigma_{jjk,1}, ..., \sigma_{jjk,N}\}$ is a diagonal matrix with the generalized eigenvalues sorted from large to small and the columns of $\mathbf{X}_{jjk} = \mathbf{Q}^{-H}_{jjk}$ contain the generalized eigenvectors. The eigenvectors are normalized such that $\mathbf{X}^H_{jjk} {\mathbf{R}}^{\text{all}}_{j} \mathbf{X}_{jjk} = \mathbf{I}$.
The optimal rank $R$ estimate $\mathbf{R}_{jjk}$ is given by keeping only the rank-1 terms belonging to the $R$ largest eigenvalues greater than one in \eqref{cemMIMO:eq:R_estimate}
\begin{equation}
\label{cemMIMO:eq:R_estimate_gevd}
p_{jk} \mathbf{R}_{jjk} = \sum_{r=1}^{R}  \frac{\sigma_{jjk,r}-1}{\tau_p-1} \mathbf{q}_{jjk,r} \mathbf{q}^H_{jjk,r} = \mathds{Q}_{jjk} \mathbb{\Lambda}_{jjk} \mathds{Q}^H_{jjk}
\end{equation}
where  $\mathbb{\Lambda}_{jjk} = \text{diag}\{\frac{\sigma_{jjk,1} -1}{\tau_p-1}, ..., $ $ \frac{\sigma_{jjk,R} -1}{\tau_p-1}\}$. 

As stated in \cite{Serizel2014}, the GEVD-based approximation effectively selects the modes with the highest SINR, since approximation errors are weighted relative to the interference and noise aggregated in ${\mathbf{R}}^{\text{all}}_{j}$. Also note that unlike other rank-approximations of \eqref{cemMIMO:eq:R_estimate}, the GEVD-based approximation is immune to scaling and linear combining of the signals, i.e. the output signal and output SINR is independent of such scaling and combining, which is a desirable property.

The proposed covariance matrix estimator is data-driven and requires only synchronization and knowledge of the chosen pilot sequence $b^t_{jk}$ of the UEs in cell $j$ in each coherence block. There is no need for prior knowledge of $\overline{\mathbf{R}}_{jjk}$ (except for the rank $R$ of $\overline{\mathbf{R}}_{jjk}$), $\overline{\mathbf{R}}_{\mathbf{n}_j\mathbf{n}_j}$, or any prior knowledge on the statistics or pilot sequences of the UEs in other cells $l \ne j$.

\subsection{Low-rank approximate MMSE estimator}
\label{cemMIMO:subsect:channel_estimation}
When the estimated quantities are used to calculate the filter in \eqref{cemMIMO:eq:W_k-optimal}, the following filter is obtained:
\begin{align}
\mathbf{W}_{jk} &= \sqrt{p_{jk}} \left( \mathbf{Q}_{jjk} \bm{\Sigma}_{jjk} \mathbf{Q}^H_{jjk} \right)^{-1} \frac{1}{p_{jk}} \mathds{Q}_{jjk} \mathbb{\Lambda}_{jjk} \mathds{Q}^H_{jjk} \\
&= \frac{1}{\sqrt{p_{jk}}} \mathds{X}_{jjk} \mathbb{\Upsilon}_{jjk} \mathds{Q}^H_{jjk} \nonumber
\end{align}
with $\mathbb{\Upsilon}_{jjk} = \text{diag}\{\upsilon_{jjk,1}, ...,  \upsilon_{jjk,R}\} =\frac{1}{\tau_p-1} \text{diag}\{\frac{\sigma_{jjk,1} - 1}{\sigma_{jjk,1}},...,$ $   \frac{\sigma_{jjk,R} - 1}{\sigma_{jjk,R}}\}$ and $\mathds{X}_{jjk} = [\mathbf{x}_{jjk,1} \ ... \ \mathbf{x}_{jjk,R}]$ contains the first $R$ columns of $\mathbf{X}_{jjk}$. The final low-rank approximate MMSE estimate for the channel of UE $k$ in coherence block $t$ is then given as
\begin{align}
\label{cemMIMO:eq:h_estimate}
\hat{\mathbf{h}}^t_{jjk} &= \mathbf{W}^H_{jk} \mathbf{y}_{j}^{t,\text{pilot}}[b^t_{jk}]
= \frac{1}{\sqrt{p_{jk}}} \mathds{Q}_{jjk} \mathbb{\Upsilon}_{jjk} \mathds{X}^H_{jjk} \mathbf{y}_{j}^{t,\text{pilot}}[b^t_{jk}] \nonumber \\
&= \frac{1}{\sqrt{p_{jk}}} \sum_{r=1}^{R} \mathbf{q}_{jjk,r} \underbrace{\upsilon_{jjk,r} \left( \mathbf{x}^H_{jjk,r} \mathbf{y}_{j}^{t,\text{pilot}}[b^t_{jk}] \right)}_{z^t_{jjk,r}}. 
\end{align}
Comparing this expression for $\hat{\mathbf{h}}^t_{jjk}$ with the model used for ${\mathbf{h}}^t_{jjk}$ in \eqref{cemMIMO:eq:model_h}, it can be observed that the estimation method produces an estimate that exactly fits in this model. The random variable $\bar{\mathbf{z}}^t_{jjk}$ of \eqref{cemMIMO:eq:model_h} is estimated from the despread signal $\mathbf{y}_{j}^{t,\text{pilot}}[b^t_{jk}]$ using a simple inner-product with the corresponding eigenvector $\mathbf{x}_{jjk,r}$ and some scalars depending on the generalized eigenvalues.

\subsection{Improved low-rank approximate MMSE estimator}
An improved low-rank approximate MMSE estimator can be derived based on \eqref{cemMIMO:eq:y}. When computing the expectation in \eqref{cemMIMO:eq:W_k-optimal}, the activity of all the UEs is represented by the random variable $\delta^{t}_{jk,li}$. Since BS $j$ is aware of the used pilot sequences of the UEs in its cell, this information can be used to improve the MMSE estimator:
\begin{align}
\label{cemMIMO:eq:y_2}
\mathbf{y}_{j}^{t,\text{pilot,impr}}[b^t_{jk}] =& \sum_{i=1}^K \bar{\delta}^{t}_{jk,ji} \sqrt{p_{ji}} \tau_p \mathbf{h}^t_{jji} \\
&+ \sum_{l \ne j} \sum_{i=1}^K \delta^{t}_{jk,li} \sqrt{p_{li}} \tau_p \mathbf{h}^t_{jli} + \mathbf{n}_{j}^{t,\text{pilot}}[b^t_{jk}] \nonumber
\end{align}
where $\bar{\delta}^{t}_{jk,ji}$ is now deterministic and either equal to 0 or 1 and $\bar{\delta}^{t}_{jk,jk} = 1$. The corresponding cell-specific covariance matrix that now depends on the coherence block index $t$ is given as (alternative to \eqref{cemMIMO:eq:R_k_dot})
\begin{align}
\label{cemMIMO:eq:R_k_dot_2}
\overline{\mathbf{R}}^{\text{pilot,impr,t}}_{jjk} &= \frac{1}{\tau_p} E \{\mathbf{y}_{j}^{t,\text{pilot,impr}}[b^t_{jk}] \mathbf{y}_{j}^{t,\text{pilot,impr},H}[b^t_{jk}] \}  \\
&= \sum_{i=1}^K \bar{\delta}^{t^2}_{jk,ji} p_{ji} \tau_p \overline{\mathbf{R}}_{jji} + \sum_{l\ne j} \sum_{i=1}^K p_{li} \overline{\mathbf{R}}_{jli} + \overline{\mathbf{R}}_{\mathbf{n}_{j}\mathbf{n}_{j}} \nonumber \\
&= \overline{\mathbf{R}}^{\text{pilot}}_{jjk} + \sum_{i \ne k} (\bar{\delta}^{t^2}_{jk,ji} \tau_p - 1) p_{ji} \overline{\mathbf{R}}_{jji}. \nonumber
\end{align}
where $ \overline{\mathbf{R}}^{\text{pilot}}_{jjk}$ is defined as in \eqref{cemMIMO:eq:R_k_dot}. By replacing the covariance matrices $\overline{\mathbf{R}}^{\text{pilot}}_{jjk}$ and $\overline{\mathbf{R}}_{jji}$ with their estimated quantities as explained in Subsection~\ref{cemMIMO:subsect:covariance_estimation}, an improved low-rank approximate MMSE estimate 
$\hat{\mathbf{h}}^{t,\text{impr}}_{jjk} = \mathbf{W}^{\text{impr},t,H}_{jk} \mathbf{y}_{j}^{t,\text{pilot,impr}}[b^t_{jk}] $ can be obtained where
\begin{equation}
    \label{cemMIMO:eq:W_2}
    \mathbf{W}^{\text{impr},t}_{jk} = \Big( \mathbf{R}^{\text{pilot}}_{jjk} + \sum_{i \ne k} (\bar{\delta}^{t^2}_{jk,ji} \tau_p - 1) p_{ji} \mathbf{R}_{jji} \Big)^{-1} \mathbf{R}_{jjk}.
\end{equation}

Although this estimator uses more of the information available at BS $j$, it is important to note that  the provided improved low-rank approximate MMSE estimator requires an extra matrix inversion for each UE in its cell, while this is not required for the low-rank approximate MMSE estimator of Subsection~\ref{cemMIMO:subsect:channel_estimation}.

\section{Numerical Simulations}
\label{cemMIMO:sect:sim}
The same multicell setup with $L = 7$ hexagonal cells as in \cite{Bjornson2017a, Neumann2018} is considered. There are $K = 10$ UEs per cell in a ring around the BS (there are 70 UEs in total) and each BS has a uniform linear array with $N = 100$ antennas. The multipath components from a UE arrive uniformly distributed in an angular interval ($10^\circ$) centered around the geographical angle to the UE \cite{Adhikary2013}. This results in channel covariance matrices with approximately 25 dominant eigenvalues.

The normalized MSE (NMSE) (averaged over 20 Monte-Carlo runs), $E\{ ||\mathbf{h}_{jjk} - \hat{\mathbf{h}}_{jjk}||^2\}/\text{tr}\{\overline{\mathbf{R}}_{jjk}\}$, for an average UE in the center cell is used as a performance metric. The proposed low-rank approximate MMSE estimator is calculated for different ranks: $R=30$ and $R=60$, denoted with \textit{GEVD 30} and \textit{GEVD 60} respectively, and also when the subtraction in \eqref{cemMIMO:eq:R_estimate} is used instead of the GEVD to estimate the channel covariance matrices, denoted with \textit{SUBT}.  The performance of the MMSE estimator with perfect channel covariance matrix knowledge (\textit{MMSE random}) in \eqref{cemMIMO:eq:W} is also shown. Similar results are shown for the proposed improved approximate MMSE estimator, denoted with the word \textit{impr}. As baseline, the simple least-square estimation (\textit{LS fixed}) that does not require covariance matrix information is shown in a scenario where each UE is allocated with a fixed pilot sequence for all coherence blocks \cite{Bjornson2017a}. Also the MMSE estimate with perfect channel covariance matrix knowledge (\textit{MMSE fixed}) for the fixed pilot allocation regime is provided as absolute lower bound.

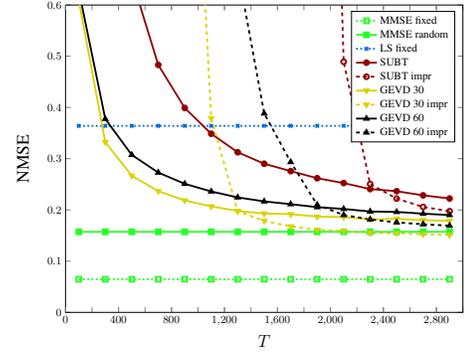
\begin{figure}[]
    \centering
    \scalebox{0.46}{
%
%
%
\begin{tikzpicture}

\begin{axis}[%
width=4.521in,
height=3.5in,
at={(0.758in,0.481in)},
scale only axis,
xmin=0,
xmax=3000,
xlabel={{\Large $T$}},
xticklabels={},
extra x ticks={0,400,800,1200,1600,2000,2400,2800},
ymin=0,
ymax=0.6,
ylabel={{\Large NMSE}},
legend style={legend cell align=left, align=left}
]
\addplot [color=mycolor1, dotted, line width=1.5pt, mark=square, mark options={solid, mycolor1}]
  table[row sep=crcr]{%
100	0.0646879383549575\\
300	0.0646879383549575\\
500	0.0646879383549575\\
700	0.0646879383549575\\
900	0.0646879383549575\\
1100	0.0646879383549575\\
1300	0.0646879383549575\\
1500	0.0646879383549575\\
1700	0.0646879383549575\\
1900	0.0646879383549575\\
2100	0.0646879383549575\\
2300	0.0646879383549575\\
2500	0.0646879383549575\\
2700	0.0646879383549575\\
2900	0.0646879383549575\\
};
\addlegendentry{MMSE fixed}

\addplot [color=mycolor1, line width=1.5pt, mark=square*, mark options={solid, fill=mycolor1, mycolor1}]
  table[row sep=crcr]{%
100	0.157213405034106\\
300	0.157213405034106\\
500	0.157213405034106\\
700	0.157213405034106\\
900	0.157213405034106\\
1100	0.157213405034106\\
1300	0.157213405034106\\
1500	0.157213405034106\\
1700	0.157213405034106\\
1900	0.157213405034106\\
2100	0.157213405034106\\
2300	0.157213405034106\\
2500	0.157213405034106\\
2700	0.157213405034106\\
2900	0.157213405034106\\
};
\addlegendentry{MMSE random}

\addplot [color=mycolor2, dotted, line width=1.5pt, mark=x, mark options={solid, mycolor2}]
  table[row sep=crcr]{%
100	0.36412706365888\\
300	0.36412706365888\\
500	0.36412706365888\\
700	0.36412706365888\\
900	0.36412706365888\\
1100	0.36412706365888\\
1300	0.36412706365888\\
1500	0.36412706365888\\
1700	0.36412706365888\\
1900	0.36412706365888\\
2100	0.36412706365888\\
2300	0.36412706365888\\
2500	0.36412706365888\\
2700	0.36412706365888\\
2900	0.36412706365888\\
};
\addlegendentry{LS fixed}

\addplot [color=mycolor3, line width=1.5pt, mark=*, mark options={solid, fill=mycolor3, mycolor3}]
  table[row sep=crcr]{%
100	NaN\\
300	1.18371728113594\\
500	0.651469996041824\\
700	0.483151283630091\\
900	0.399061139771051\\
1100	0.348746747655379\\
1300	0.312600088609903\\
1500	0.290157766267836\\
1700	0.275785452564284\\
1900	0.261917969752354\\
2100	0.252419840943378\\
2300	0.240754323989333\\
2500	0.236532636330284\\
2700	0.228472434367609\\
2900	0.222345980093549\\
};
\addlegendentry{SUBT}

\addplot [color=mycolor3, dashed, line width=1.5pt, mark=o, mark options={solid, mycolor3}]
  table[row sep=crcr]{%
100	NaN\\
300	NaN\\
500	NaN\\
700	NaN\\
900	NaN\\
1100	NaN\\
1300	NaN\\
1500	NaN\\
1700	9.10973822831291\\
1900	2.16928450124048\\
2100	0.489218054723536\\
2300	0.250561238153045\\
2500	0.222093607484023\\
2700	0.205830263187746\\
2900	0.197505973146327\\
};
\addlegendentry{SUBT impr}

\addplot [color=mycolor4, line width=1.5pt, mark=triangle*, mark options={solid, rotate=180, fill=mycolor4, mycolor4}]
  table[row sep=crcr]{%
100	0.609035814824853\\
300	0.332435670735282\\
500	0.266733460540579\\
700	0.236600969530268\\
900	0.218610074490991\\
1100	0.206971253708324\\
1300	0.197913741957602\\
1500	0.192893339868004\\
1700	0.191716643112951\\
1900	0.186953868238725\\
2100	0.186018263472102\\
2300	0.181630311611402\\
2500	0.18254088009559\\
2700	0.180009942346127\\
2900	0.17874672252259\\
};
\addlegendentry{GEVD 30}

\addplot [color=mycolor4, dashed, line width=1.5pt, mark=triangle, mark options={solid, rotate=180, mycolor4}]
  table[row sep=crcr]{%
100	NaN\\
300	NaN\\
500	11.7030112012024\\
700	11.6760090741305\\
900	1.35913364434291\\
1100	0.378779367558433\\
1300	0.19752698489915\\
1500	0.178600967020157\\
1700	0.16788045374692\\
1900	0.161205400849231\\
2100	0.158785434916248\\
2300	0.155182253172881\\
2500	0.154292219296427\\
2700	0.152536462852771\\
2900	0.151084481756252\\
};
\addlegendentry{GEVD 30 impr}

\addplot [color=black, line width=1.5pt, mark=triangle*, mark options={solid, fill=black, black}]
  table[row sep=crcr]{%
100	0.628918605886917\\
300	0.37772272318894\\
500	0.307397109083268\\
700	0.272949675008532\\
900	0.251002481121807\\
1100	0.236015356970047\\
1300	0.224440718260702\\
1500	0.216651385537535\\
1700	0.21121274405504\\
1900	0.205397045904979\\
2100	0.20190524625122\\
2300	0.196888471374513\\
2500	0.196084570311415\\
2700	0.192696830287373\\
2900	0.189998548464053\\
};
\addlegendentry{GEVD 60}

\addplot [color=black, dashed, line width=1.5pt, mark=triangle, mark options={solid, black}]
  table[row sep=crcr]{%
100	NaN\\
300	NaN\\
500	NaN\\
700	6.21073134332681\\
900	4.13634698184654\\
1100	1.47656215364177\\
1300	0.669026583437771\\
1500	0.388285652470005\\
1700	0.293478478920974\\
1900	0.209960174172488\\
2100	0.190103477413587\\
2300	0.181403135434654\\
2500	0.175591085513578\\
2700	0.172000543218172\\
2900	0.169232464722833\\
};
\addlegendentry{GEVD 60 impr}

\end{axis}
\end{tikzpicture}%
}  
  
    \caption{NMSE for an average intra-cell UE for different estimators with $\tau_p=10$ when the number of coherence block is varied.}
    \label{cemMIMO:fig:T_run}
\end{figure}

Figure~\ref{cemMIMO:fig:T_run} shows that a relative small number of coherence blocks (300) suffices to obtain an accurate channel estimate for the proposed methods. The estimators with a rank constraint on the channel covariance matrix outperform the estimators without the rank constraint (SUBT) and $R=30$ works better for the current setup than $R=60$, since it can already capture the dominant eigenspace. The proposed GEVD-based procedure focuses on the modes with the highest power in the estimated covariance matrices and neglect the other modes, that typically contain more estimation error noise than useful signal power. The improved approximate MMSE estimator is also outperforming the approximate MMSE estimator when there are enough coherence blocks available for the estimation procedure. Since all the estimates of the channel covariance matrices are used to calculate the filter in \eqref{cemMIMO:eq:W_2}, this will only perform well when they are estimated using enough samples. Although not shown in Figure~\ref{cemMIMO:fig:T_run}, it should be noted that the \textit{SUBT}-curve will converge to the \textit{MMSE fixed}-curve when $T \rightarrow \infty$ while the \textit{GEVD}-curves perform worse, since the rank assumption on the channel covariance matrices is only approximately valid for this setup.

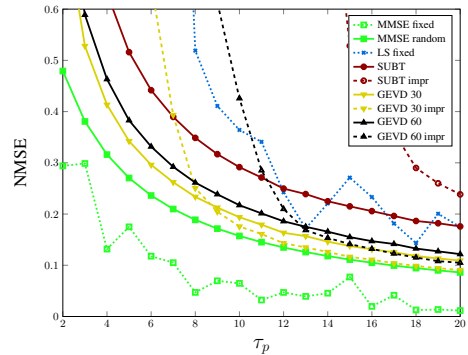
\begin{figure}[]
    \centering
    \scalebox{0.46}{
%
%
%
\begin{tikzpicture}

\begin{axis}[%
width=4.521in,
height=3.5in,
at={(0.758in,0.481in)},
scale only axis,
xmin=2,
xmax=20,
xlabel={{\LARGE $\tau_p$}},
ymin=0,
ymax=0.6,
ylabel={{\Large NMSE}},
legend style={legend cell align=left, align=left}
]
\addplot [color=mycolor1, dotted, line width=1.5pt, mark=square, mark options={solid, mycolor1}]
  table[row sep=crcr]{%
2	0.293994362040285\\
3	0.298736157819136\\
4	0.131570664604096\\
5	0.175022451665695\\
6	0.117747267699474\\
7	0.105048911899541\\
8	0.0470523331890899\\
9	0.0696428000435301\\
10	0.0646879383549575\\
11	0.0321971486172592\\
12	0.0470151589413137\\
13	0.0392518296717464\\
14	0.0455087877805241\\
15	0.077175700565107\\
16	0.019710910807855\\
17	0.0415036479909447\\
18	0.0127237496548051\\
19	0.0136521977306023\\
20	0.0116193597030164\\
};
\addlegendentry{MMSE fixed}

\addplot [color=mycolor1, line width=1.5pt, mark=square*, mark options={solid, fill=mycolor1, mycolor1}]
  table[row sep=crcr]{%
2	0.479070009006434\\
3	0.3807206633475\\
4	0.316068716928133\\
5	0.270302888676092\\
6	0.2361891707797\\
7	0.209772581202985\\
8	0.188707525876317\\
9	0.171514374310066\\
10	0.157213405034106\\
11	0.145130019500544\\
12	0.13478447779544\\
13	0.125826243559618\\
14	0.117993207925791\\
15	0.111085444442071\\
16	0.104947790906644\\
17	0.0994579775965001\\
18	0.0945183447505872\\
19	0.0900499437030575\\
20	0.0859882577611948\\
};
\addlegendentry{MMSE random}

\addplot [color=mycolor2, dotted, line width=1.5pt, mark=x, mark options={solid, mycolor2}]
  table[row sep=crcr]{%
2	NaN\\
3	3.29068244272392\\
4	2.83527338787835\\
5	1.72825412731775\\
6	1.40126326066043\\
7	1.50407922625533\\
8	0.518930267492193\\
9	0.410880938017573\\
10	0.36412706365888\\
11	0.340936647028866\\
12	0.242522662824389\\
13	0.171791904605244\\
14	0.220685975189696\\
15	0.271163102130532\\
16	0.233170433813577\\
17	0.181531666534685\\
18	0.143303144630884\\
19	0.200577829436436\\
20	0.174782557651946\\
};
\addlegendentry{LS fixed}

\addplot [color=mycolor3, line width=1.5pt, mark=*, mark options={solid, fill=mycolor3, mycolor3}]
  table[row sep=crcr]{%
2	1.34874677769883\\
3	0.834900185283727\\
4	0.631165377357453\\
5	0.516026595274089\\
6	0.441714715058534\\
7	0.389592371304524\\
8	0.348694232049931\\
9	0.31685507545769\\
10	0.29151676159286\\
11	0.271264031611707\\
12	0.249695973499823\\
13	0.238988107236926\\
14	0.224945875667806\\
15	0.214998153026805\\
16	0.205702621659012\\
17	0.196226968888936\\
18	0.186480941761496\\
19	0.18201977338918\\
20	0.175881034451251\\
};
\addlegendentry{SUBT}

\addplot [color=mycolor3, dashed, line width=1.5pt, mark=o, mark options={solid, mycolor3}]
  table[row sep=crcr]{%
2	NaN\\
3	NaN\\
4	NaN\\
5	NaN\\
6	NaN\\
7	NaN\\
8	NaN\\
9	NaN\\
10	NaN\\
11	NaN\\
12	3.91877120818741\\
13	3.13422808969192\\
14	0.91939953772038\\
15	0.528329732557155\\
16	0.385656503904767\\
17	0.36\\
18	0.29\\
19	0.26\\
20	0.238423842306582\\
};
\addlegendentry{SUBT impr}

\addplot [color=mycolor4, line width=1.5pt, mark=triangle*, mark options={solid, rotate=180, fill=mycolor4, mycolor4}]
  table[row sep=crcr]{%
2	0.793675831022847\\
3	0.528027548937989\\
4	0.413351735420683\\
5	0.342285647751875\\
6	0.296019726588645\\
7	0.262087059722062\\
8	0.233636072203405\\
9	0.212616730313768\\
10	0.193832797880717\\
11	0.179220336241342\\
12	0.163449038137145\\
13	0.156830040392248\\
14	0.146081292785794\\
15	0.137569053378755\\
16	0.131098510691417\\
17	0.125411258546501\\
18	0.118130541732097\\
19	0.113185901657472\\
20	0.109211857839952\\
};
\addlegendentry{GEVD 30}

\addplot [color=mycolor4, dashed, line width=1.5pt, mark=triangle, mark options={solid, rotate=180, mycolor4}]
  table[row sep=crcr]{%
2	NaN\\
3	NaN\\
4	8.18746971937719\\
5	2.88743247722572\\
6	0.649358696001636\\
7	0.393330409946041\\
8	0.250420203547804\\
9	0.204543846991556\\
10	0.175676493989381\\
11	0.160982234958789\\
12	0.143024915063084\\
13	0.134826480190473\\
14	0.125963376914635\\
15	0.116507522640436\\
16	0.110968635599617\\
17	0.104010579581846\\
18	0.0987681654770553\\
19	0.0943414515213388\\
20	0.0907097956245584\\
};
\addlegendentry{GEVD 30 impr}

\addplot [color=black, line width=1.5pt, mark=triangle*, mark options={solid, fill=black, black}]
  table[row sep=crcr]{%
2	0.866472341186508\\
3	0.588815128434734\\
4	0.463173301532775\\
5	0.382861038015337\\
6	0.331632894730235\\
7	0.292184917415826\\
8	0.261337359226084\\
9	0.238530848503092\\
10	0.217271664858311\\
11	0.201295808687334\\
12	0.185875791739564\\
13	0.175186611626417\\
14	0.165602351723277\\
15	0.154987312475604\\
16	0.147064345220793\\
17	0.141563703781286\\
18	0.132617869033661\\
19	0.127182352841998\\
20	0.121726892012043\\
};
\addlegendentry{GEVD 60}

\addplot [color=black, dashed, line width=1.5pt, mark=triangle, mark options={solid, black}]
  table[row sep=crcr]{%
2	NaN\\
3	NaN\\
4	NaN\\
5	NaN\\
6	NaN\\
7	3.09964965769327\\
8	1.15107306693873\\
9	0.612917660608512\\
10	0.425835362427303\\
11	0.285161345955478\\
12	0.209824746828535\\
13	0.168590072811406\\
14	0.153562550794649\\
15	0.141298937971992\\
16	0.13197534549247\\
17	0.122788925995106\\
18	0.115116308869014\\
19	0.108022808459123\\
20	0.104916611257487\\
};
\addlegendentry{GEVD 60 impr}

\end{axis}
\end{tikzpicture}%
}  
  
    \caption{NMSE for an average intra-cell UE for different estimators with $T=1500$ when the number of pilot sequences is varied.}
    \label{cemMIMO:fig:tau_run}
\end{figure}

In Figure~\ref{cemMIMO:fig:tau_run}, the performance with respect to the number of available pilot sequences $\tau_p$ is shown for $T=1500$ coherence blocks. The seemingly arbitrary shapes of the curves with a fixed pilot allocation emerge from the fact that the pilots are allocated in a cyclic way, starting from the leftmost UE in a cell and the cycle is not restarted in the neighboring cell. It is again observed that the proposed low-rank estimators outperform the estimates obtained with subtraction of the covariance matrices and the baseline LS-method. \textit{GEVD 30} provides the best performance, independently of the number of available pilot sequences. Increasing $\tau_p$, increases the dominance of $\overline{\mathbf{R}}_{jjk}$ in \eqref{cemMIMO:eq:R_k_dot}, and has thus a positive impact on the channel estimation performance.

\section{Conclusion}
\label{cemMIMO:sect:con}
In this paper, a new uplink channel covariance matrix estimator has been presented for low-rank channel covariance matrices, using a GEVD of two covariance matrices that are estimated from the available uplink data. The requirements for the systems are minimal, i.e. UEs in a cell can randomly choose a pilot sequence in each coherence block and only the BS in its cell should be aware of its choice. Except for synchronization, there is no need for communication between the different BSs and no prior knowledge on the background noise is required. The derived approximate MMSE and improved approximate MMSE estimators have been shown to provide a good estimate of the true channel, even when only a small number of coherence blocks is available, while a simple subtraction of the two covariance matrices provides inferior performance due to estimation errors.

\bibliographystyle{IEEEbib}

\end{document}